\documentclass{article}
\usepackage{spconf,amsmath,amssymb,graphicx}
\usepackage{cite,booktabs,multirow,xcolor,flafter}
\usepackage{enumitem}
\usepackage[hidelinks]{hyperref}
\usepackage{microtype}
\title{SEAL: Mixture-Closed Additive Reconstruction and Refinement-Aware Expert Routing for Efficient Speech Separation}
\name{Shao-Chun Hu$^{1}$, Zi-Xiang Lin$^{1}$, Jeih-Weih Hung$^{1}$, Hung-Shin Lee$^{2}$}
\address{$^{1}$Department of Electrical Engineering, National Chi Nan University\\ $^{2}$Graduate Institute of AI Interdisciplinary Applied Technology, National Taiwan Normal University}
\begin{document}
\ninept
\maketitle

\begin{abstract}
Compact time-frequency separators that mask the mixture and refine through a shared cell face two limits. First, a bounded multiplicative mask only scales a mixture bin, so where overlapping components cancel, the estimate stays small. Second, a shared cell applies the same weights to every time-frequency token at every step, so enlarging it adds compute everywhere. We present SEAL (Sparse Expert routing with Additive Latent reconstruction) to address both. For reconstruction, a zero-sum additive residual bounded by the local mixture amplitude lets estimates be nonzero where components cancel yet still sum to the mixture. For routing, a query built from acoustic and inter-step evidence sends each token to one of six residual experts, and a norm cap keeps the step cue from overriding clear acoustic evidence. On EchoSet, SEAL (small) surpasses TIGER (small) by 0.31~dB SI-SDRi with 28\% fewer parameters and 2.9 times fewer MACs, and SEAL (large) is within 0.07~dB SI-SDRi of TIGER (large) at 3.1 times fewer MACs.
\end{abstract}


\section{INTRODUCTION}

Single-channel speech separation has moved from discriminative embeddings and permutation-invariant training~\cite{hershey2016dc,yu2017pit,kolbaek2017upit} to time-domain and dual-path models~\cite{convtasnet,dprnn,subakan2021sepformer}, and multiscale and group-communication designs have cut its cost~\cite{sudormrf,afrcnn,tdanet,luo2021groupcomm}. On EchoSet, the noisy reverberant corpus it introduced, TIGER~\cite{xu2025tiger} sets the current efficiency reference.

Despite this progress, two limits persist for a compact time--frequency separator that masks the mixture and refines through a shared cell. First, a bounded multiplicative mask limits its output where overlapping components cancel, because the estimate is a scaling of a mixture bin that destructive interference has already made small. Second, a shared iterative separator applies the same weights to every time--frequency token at every step, although a token's current acoustic state, its change since the previous step, and the step index carry different information for choosing a correction; enlarging the cell to serve them all adds compute everywhere. To address both limits, we present SEAL (\textbf{S}parse \textbf{E}xpert routing with \textbf{A}dditive \textbf{L}atent reconstruction)\footnote{Code is available at \href{https://github.com/Xiang-Lin75/SEAL}{https://github.com/Xiang-Lin75/SEAL}.}, a shared iterative separator.

For the first limit, complex masking and spectral mapping reconstruct more freely~\cite{williamson2016crm,wang2023tfgrid,zhao2023d2former}, but a mapped estimate carries no bound tied to the observation. Mixture consistency constrains the estimates to explain the observation~\cite{wisdom2019consistency}, and noise estimation can support separation in noisy mixtures~\cite{zhang2024nass}. SEAL keeps an observation-tied bound and the mixture sum while relaxing the per-bin limit: a mixture-closed allocation is combined with an additive residual bounded by the local mixture amplitude, so estimates can be nonzero at cancelled bins, and a residual sink absorbs what the speech groups do not claim.

For the second limit, sparse mixture-of-experts layers select a function per token through conditional execution, increasing stored capacity~\cite{shazeer2017moe,fedus2022switch}. In speech separation, they have addressed the quality--compute and overlap--artifact trade-offs~\cite{wang2022moess} and replaced the feed-forward modules of stacked time- and frequency-wise Conformer blocks~\cite{hu2026tfmoe}; shared layers and recurrent routers extend them further~\cite{csordas2024moeut,qiu2025rmoe}. In a shared iterative separator, the step index also enters the router, and an unconstrained step cue added to the routing query can override acoustic evidence: in our control without the norm cap it reaches 1.7--1.9 times the mean norm of the fused evidence at the first two steps, and removing it changes 85.5\% of first-step assignments (Section~\ref{sec:cue-results}). SEAL instead keeps the inter-step change as explicit evidence in the routing query and bounds the norm of the step cue.

SEAL couples both mechanisms in one shared cell. Our contributions are:
\begin{enumerate}[nosep,leftmargin=*,label=\arabic*.]
\item \textbf{Mixture-closed additive reconstruction.} Complex masks built from latent occupancy and group ownership are combined with a zero-sum additive residual bounded by the local mixture amplitude, so the estimates can be nonzero at cancelled bins and still sum to the mixture.
\item \textbf{Refinement-aware bounded routing.} A query fused from up to five evidence branches, including the inter-step delta and within-step temporal statistics, routes each token to one of six stateless residual experts, and the step cue is capped at norm $\rho$ so that it cannot flip any routing decision whose acoustic Top-2 margin exceeds $2\rho$.
\item \textbf{Cost--quality operating points on EchoSet.} SEAL (small) exceeds TIGER (small) at 2.9 times fewer MACs and lower measured latency, SEAL (large) matches the SI-SDRi of TIGER (large) to within 0.07~dB at 3.1 times fewer MACs, and retrained controls show that removing either mechanism drops SEAL (small) below TIGER (small) (Table~\ref{tab:ablation}).
\end{enumerate}

\section{PROPOSED METHODOLOGY}

\begin{figure*}[t]
\centering
\includegraphics[width=\textwidth]{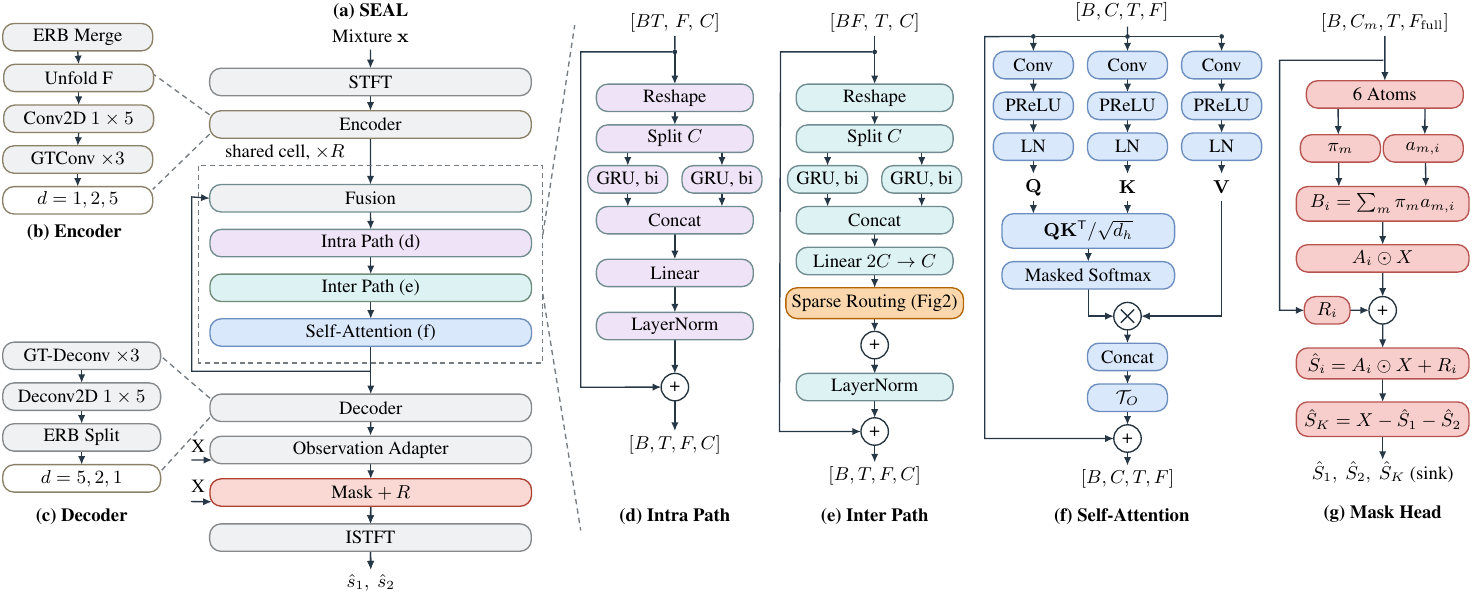}
\vspace{-20pt}
\caption{SEAL architecture. In (f), $d_h=130$ is the query dimension and $\mathcal{T}_O$ the $1\times1$ convolution after head concatenation. In (g), $A_i$ also includes the correction \eqref{eq:mask-correction}, $C_m=24$ is the head width, and $F_{\rm full}$ the full frequency resolution. Arrows labeled $X$ carry the mixture STFT.}
\label{fig:arch}
\end{figure*}

\subsection{Shared Refinement Backbone}

Let the mixture $\mathbf x=\sum_{i=1}^{N_s}\mathbf s_i+\mathbf n$ contain $N_s=2$ speech signals and background $\mathbf n$. Its STFT $\mathbf X\in\mathbb C^{T\times F_0}$ has $T$ frames and $F_0=257$ bins, using a 512-sample window and 256-sample hop at 16~kHz. The output has $K=N_s+1$ groups: $N_s$ speakers and a residual sink.

We adapt GTCRN~\cite{rong2024gtcrn} (grouped temporal convolutions and grouped dual-path GRUs) as the backbone with bidirectional inter-frame GRUs, keeping its equivalent rectangular bandwidth compression and downsampling to an anchor $\mathbf a\in\mathbb R^{C\times T\times F}$ with $C=72$ channels and $F=65$ frequencies (Fig.~\ref{fig:arch}(b)--(f)). The temporal self-attention in Fig.~\ref{fig:arch}(f), an implementation choice, replaces GTCRN's recurrent channel attention; Section~\ref{sec:results} reports its cost beyond 16~s. A single cell shares its parameters $\boldsymbol\theta$ across $R=4$ latent refinement steps~\cite{bralios2023latent}:
\begin{equation}
\begin{aligned}
&\mathbf u_1=\mathbf a,\quad
\mathbf u_s=F_s(\mathbf a,\mathbf z_{s-1})\quad(s>1),\\
&(\mathbf z_s,\mathbf H_s)=G_{\boldsymbol\theta}(\mathbf u_s,\mathbf a;s).
\end{aligned}
\label{eq:refine}
\end{equation}
Here $G_{\boldsymbol\theta}$ returns a state $\mathbf z_s$ and a readout $\mathbf H_s$, and $F_s$ fuses $\mathbf a$ with $\mathbf z_{s-1}$ through step-dependent embeddings. GRU hidden states are re-initialized at every step, so $\mathbf z_{s-1}$ and the readout delta of Section~\ref{sec:cue} are the only cross-step paths. After the last refinement step, the decoder restores the full frequency resolution, and a decoder-gated observation adapter re-injects $[\log(1+|X|),\,r\Re X/|X|,\,r\Im X/|X|]$ before the $C_m=24$ channel head, where $r=|X|/(|X|+\delta)$ down-weights the phase of low-magnitude bins and $\delta$ is 0.01 times the mean mixture magnitude of the utterance.

\subsection{Mixture-Closed Additive Reconstruction}
\label{sec:recon}

At each bin, the head predicts occupancy logits $\mathbf o\in\mathbb R^M$ for $M=6$ atoms and ownership logits $\mathbf v_m\in\mathbb R^K$ for each atom. We omit $(t,f)$ below, so $X$, $A_k$, and $R_k$ denote scalar coefficients. With $\pi_m=[\operatorname{softmax}(\mathbf o)]_m$ and $a_{mk}=[\operatorname{softmax}(\mathbf v_m)]_k$, define
\begingroup
\setlength{\abovedisplayskip}{2pt plus 1pt minus 1pt}
\setlength{\belowdisplayskip}{2pt plus 1pt minus 1pt}
\setlength{\abovedisplayshortskip}{0pt plus 1pt}
\setlength{\belowdisplayshortskip}{1pt plus 1pt minus 1pt}
\begin{equation}
B_k=\sum_{m=1}^{M}\pi_m a_{mk},\qquad
A_k=\sum_{m=1}^{M}\pi_m(a_{mk}+c_{mk}),
\label{eq:allocation}
\end{equation}
\endgroup
where $k$ indexes all $K$ groups, $i\le N_s$ the speech groups only, $m$ the atoms, and $c_{mk}$ is a zero-sum complex correction \eqref{eq:mask-correction}. $B_k$ is the resulting real allocation (summing to one across $K$), and $A_k$ the complex mask. Real-valued logits $\xi^{\Re}_{mk},\xi^{\Im}_{mk}$ construct the correction:
\begin{equation}
\begin{aligned}
v^c_{mk}&=\gamma a_{mk}(\tanh\xi^{\Re}_{mk}+\mathrm j\tanh\xi^{\Im}_{mk}),\\
c^0_{mk}&=v^c_{mk}-K^{-1}\sum_{\ell=1}^{K}v^c_{m\ell},\\
c_{mk}&=\frac{c^0_{mk}}{\max(1,\max_\ell|c^0_{m\ell}|/\gamma)},\qquad\gamma=0.1,
\end{aligned}
\label{eq:mask-correction}
\end{equation}
where $\mathrm j$ is the imaginary unit and $\ell$ indexes groups inside the sums. The $\tanh$ bounds the real and imaginary parts of $v^c_{mk}$ by $\gamma a_{mk}$, so a correction scales with the group's share of the atom; centering removes the mean over the $K$ groups, and the denominator rescales all $K$ corrections of atom $m$ together when the largest exceeds $\gamma=0.1$. Hence $\sum_k c_{mk}=0$, $|c_{mk}|\le\gamma$, and $\sum_k A_k=1$. Atoms factorize the allocation without assigning physical source identities.

The local mixture amplitude $L$ is the square root of the $3\times3$ time--frequency average-pooled mixture power at bin $(t,f)$. With a learned scalar $\eta$ and the logistic sigmoid $\sigma$, we set $\alpha=0.5\sigma(\eta)$. Real-valued outputs $h_i^{\Re},h_i^{\Im}$ give the speech proposals and a balancing sink proposal:
\begingroup
\setlength{\abovedisplayskip}{2pt plus 1pt minus 1pt}
\setlength{\belowdisplayskip}{2pt plus 1pt minus 1pt}
\setlength{\abovedisplayshortskip}{0pt plus 1pt}
\setlength{\belowdisplayshortskip}{1pt plus 1pt minus 1pt}
\begin{equation}
u_i=\alpha L\sqrt{B_i}(\tanh h_i^{\Re}+\mathrm j\tanh h_i^{\Im}),\quad
u_K=-\sum_{i=1}^{N_s}u_i,
\label{eq:proposal}
\end{equation}
\endgroup
where the proposal magnitude scales as $\sqrt{B_i}$ against $L$ and a learned global scale $\alpha\in(0,0.5)$, and $u_K$ ensures $\sum_{k=1}^K u_k=0$. All groups share one denominator:
\begin{equation}
Z_R=\max\!\left(1,\frac{\max_{k\le K}|u_k|}{\alpha L}\right),\;
R_k=\frac{u_k}{Z_R},\; \widehat S_k=A_kX+R_k,
\label{eq:recon}
\end{equation}
where $Z_R\ge1$ shrinks all $K$ proposals by one factor whenever a magnitude exceeds the budget $\alpha L$; for $\alpha L=0$, set $R_k=0$. Hence $\sum_{k=1}^{K}R_k=0$ by \eqref{eq:proposal}, $|R_k|\le0.5L$ by $Z_R$ and $\alpha\le0.5$, and $\sum_{k=1}^{K}\widehat S_k=X$ by $\sum_kA_k=1$. The sink, recomputed as $X-\sum_{i=1}^{N_s}\widehat S_i$, is not supervised as a noise estimate: it absorbs whatever the speech groups do not claim, so the zero-sum constraint does not couple the speech proposals, and only the speech outputs undergo the inverse STFT.

Since $|A_k|\le1+\gamma$, the mask term obeys $|A_kX|\le1.1|X|$, whereas at a bin with $X=0$ but $L>0$ the additive residual still permits a nonzero estimate. The branch predicting $h_i^{\Re},h_i^{\Im}$ is a $1\times1$--depthwise--$1\times1$ convolutional head with 990 parameters whose output layer is zero-initialized.

\subsection{Sparse Residual Routing}
\label{sec:routing}

For a time--frequency token, let $\mathbf r_s\in\mathbb R^{2C}$ be the bidirectional grouped inter-frame GRUs' output at step $s$ (Fig.~\ref{fig:arch}(e)). Its shared readout is $\mathbf H_s=\mathbf W_{\rm sh}\mathbf r_s+\mathbf b_{\rm sh}\in\mathbb R^C$, with weights shared across steps. The two GRU outputs are layer-normalized separately into $\bar{\mathbf r}_s\in\mathbb R^{2C}$. Each of $E=6$ stateless experts $\mathcal E_j:\mathbb R^{2C}\to\mathbb R^C$ applies a width-96 SiLU value projection gated by a sigmoid to both GRU outputs, projecting their concatenation to $C$ channels. Given routing probabilities $p_{s,j}$ (Section~\ref{sec:cue}), the corrected readout is
\begingroup
\setlength{\abovedisplayskip}{2pt plus 1pt minus 1pt}
\setlength{\belowdisplayskip}{2pt plus 1pt minus 1pt}
\setlength{\abovedisplayshortskip}{0pt plus 1pt}
\setlength{\belowdisplayshortskip}{1pt plus 1pt minus 1pt}
\begin{equation}
j^*=\arg\max_j p_{s,j},\qquad
\widetilde{\mathbf H}_s=\mathbf H_s+
\beta_s\mathcal E_{j^*}(\bar{\mathbf r}_s).
\label{eq:readout}
\end{equation}
\endgroup
The corrected readout $\widetilde{\mathbf H}_s$ replaces $\mathbf H_s$ in the remainder of the cell (LayerNorm, residual sum, and attention in Fig.~\ref{fig:arch}(e)--(f)) whereas the uncorrected $\mathbf H_s$ is passed to the next step, where it forms the inter-step delta. The correction strength $\beta_s=0.5\sigma(\cdot)$ is predicted by a separate multilayer perceptron from the routing query and summaries of $\mathbf r_s$, $\mathbf H_s$ and the delta, not from the routing probability $p_{s,j^*}$.

\subsection{Refinement-Aware Query and Bounded Cue}
\label{sec:cue}

\begin{figure}[t]
\centering
\includegraphics[width=1\columnwidth]{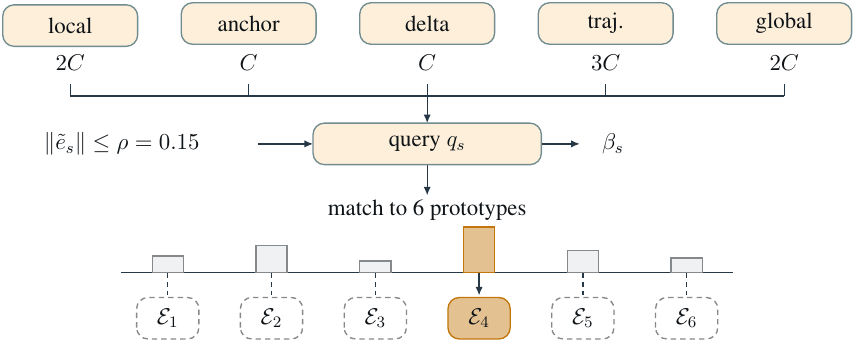}
\vspace{-20pt}
\caption{Refinement-aware Top-1 residual routing. ``Traj.'' denotes within-step temporal statistics; the score profile is schematic.}
\label{fig:readout}
\end{figure}

Five evidence branches $b\in\{1,\dots,5\}$ (Fig.~\ref{fig:readout}) project their inputs to $L_2$-normalized vectors $\mathbf f_{s,b}\in\mathbb R^{24}$, each through layer normalization and a linear map. Their inputs are the local recurrent features $\mathbf r_s$ ($2C$), the anchor ($C$), the inter-step delta ($C$), the temporal statistics ($3C$), and the global context ($2C$). The delta is $\Delta\mathbf H_s=\mathbf H_s-\operatorname{stopgrad}(\mathbf H_{s-1})$, disabled at step one; $\operatorname{stopgrad}$ is the identity with zero gradient. The temporal statistics are the per-frequency time mean, standard deviation, and mean absolute first difference of $\mathbf H_s$, broadcast over time, and the global context is the mean and standard deviation of the anchor over valid positions, shared across tokens.

A softmax over learned linear scores of each $\mathbf f_{s,b}$, plus a step-dependent bias, gives the branch weights $w_{s,b}$. With a learned additive step embedding $\mathbf e_s\in\mathbb R^{24}$ per step, define
\begingroup
\setlength{\abovedisplayskip}{2pt plus 1pt minus 1pt}
\setlength{\belowdisplayskip}{2pt plus 1pt minus 1pt}
\setlength{\abovedisplayshortskip}{0pt plus 1pt}
\setlength{\belowdisplayshortskip}{1pt plus 1pt minus 1pt}
\begin{equation}
\mathbf f_s=\sum_b w_{s,b}\mathbf f_{s,b},\qquad
\widetilde{\mathbf e}_s=
\frac{\mathbf e_s}{\max(1,\|\mathbf e_s\|_2/\rho)},\quad\rho=0.15.
\label{eq:clip}
\end{equation}
\endgroup
Here $\mathbf f_s\in\mathbb R^{24}$ is the fused acoustic evidence, a convex combination of unit vectors, and $\widetilde{\mathbf e}_s$ is the step cue projected onto the ball of radius $\rho=0.15$. The routing query is the unit-norm vector $\mathbf q_s=\operatorname{normalize}(\mathbf f_s+\widetilde{\mathbf e}_s)$, where $\operatorname{normalize}$ is $L_2$ normalization with a numerical floor, and the probability $p_{s,j}$ of expert $j$ is a softmax over the $E$ experts of the cosine similarities $\mathbf q_s^\top\bar{\mathbf P}_j/\tau$ to learned $L_2$-normalized prototypes $\bar{\mathbf P}_j\in\mathbb R^{24}$ with a learned temperature $\tau\in[0.1,2]$. Adding the cue changes the gap $g_{jk}=\mathbf f_s^\top(\bar{\mathbf P}_j-\bar{\mathbf P}_k)$ by at most $\rho\|\bar{\mathbf P}_j-\bar{\mathbf P}_k\|_2\le2\rho$, so every gap with $|g_{jk}|>2\rho$ keeps its sign; the bound is absolute rather than relative to $\|\mathbf f_s\|_2$, which can be small, and leaves the step-dependent branch biases and fusion embeddings unconstrained.

\section{EXPERIMENTAL SETUP}
\label{sec:setup}

\subsection{Dataset and Evaluation Metrics}
We used the official 20,268/4,604/2,650 training\slash validation\slash test split of EchoSet~\cite{xu2025tiger}, a 16~kHz two-speaker noisy corpus whose reverberation is simulated with SoundSpaces~2.0~\cite{chen2022soundspaces2}. We report SI-SDRi~\cite{leroux2019sdr} without mean removal~\cite{xu2025tiger}, SDRi~\cite{vincent2006bss} with a 512-tap filtered projection~\cite{scheibler2022sdr}, and MACs following TIGER~\cite{xu2025tiger}. We timed batch-1 fp32 latency for SEAL and TIGER on one host (RTX~5090; one AMD EPYC~9554 thread) as the mean, after warm-up, over 1,000 GPU and 20 CPU waveform-to-waveform forward passes at 1~s, with fewer passes for longer inputs; our TIGER GPU times are within 6\% of its published RTX~4090 times. Fixed-checkpoint diagnostics report utterance-bootstrap 95\% confidence intervals (10,000 resamples).

\subsection{Training Configuration}
We trained SEAL (small, $R=4$; Full in Table~\ref{tab:ablation}) and SEAL (large, $R=8$) with Adam (batch size 4, learning rate 0.001), maximizing zero-mean SNR under permutation-invariant training~\cite{kolbaek2017upit}. The objective adds complex-spectral and waveform $L_1$ terms (unit weight) and three auxiliary terms (weight 0.001): load-balance loss~\cite{fedus2022switch}, router z-loss~\cite{zoph2022stmoe}, and prototype orthogonality. Training ran up to 500 epochs with early stopping on validation loss (patience 20), keeping the best checkpoint; the learning rate was halved when the validation loss failed to improve by $10^{-4}$ for 10 epochs. Each reported configuration is a single training run; none is repeated across seeds. For router gradients, $\beta_s$ in \eqref{eq:readout} is multiplied by $1+p_{s,j^*}-\operatorname{stopgrad}(p_{s,j^*})$, which is one in the forward pass.

\section{RESULTS AND DISCUSSION}
\label{sec:results}

\subsection{Comparisons with State-of-the-Art Systems}

\begin{table}[t]
\caption{EchoSet test results. Baseline scores, parameters and MACs are from TIGER~\cite{xu2025tiger} (\texttt{ptflops~0.7.3}, one second of 16~kHz input); SEAL is profiled identically. Bold: best per column.}
\vspace{5pt}
\label{tab:external}
\centering
\small
\setlength{\tabcolsep}{0pt}
\begin{tabular*}{\columnwidth}{@{\extracolsep{\fill}}lrrrr@{}}
\toprule
\multirow[c]{2}{*}{Method} & SDRi & SI-SDRi & Params & MACs \\
       & (dB) & (dB)    & (M)    & (G) \\
\midrule
Conv-TasNet~\cite{convtasnet}      &  7.69 &  6.89 &  5.62 &   7.19 \\
DPRNN~\cite{dprnn}                &  5.87 &  5.06 &  2.72 &  45.01 \\
SudoRM-RF 1.0x~\cite{sudormrf}     &  7.70 &  6.84 &  2.72 &   4.65 \\
A-FRCNN-16~\cite{afrcnn}          &  9.64 &  8.76 &  6.13 &  81.28 \\
TDANet Large~\cite{tdanet}        & 10.14 &  9.21 &  2.33 &   9.19 \\
BSRNN~\cite{bsrnn}               & 12.75 & 12.23 & 25.97 &  98.70 \\
TF-GridNet~\cite{wang2023tfgrid}   & 13.73 & 12.85 & 14.43 & 323.75 \\
TIGER (small)~\cite{xu2025tiger}     & 13.15 & 12.58 &  0.82 &   7.65 \\
TIGER (large)~\cite{xu2025tiger}     & 14.22 & \textbf{13.73} & 0.82 & 15.27 \\
\midrule
SEAL (small) & 13.62 & 12.89 & \textbf{0.59} & \textbf{2.64} \\
SEAL (large) & \textbf{14.42} & 13.66 & \textbf{0.59} & 4.85 \\
\bottomrule
\end{tabular*}
\vspace{5pt}
\end{table}

\begin{figure}[t]
\centering
\includegraphics[width=1.0\linewidth]{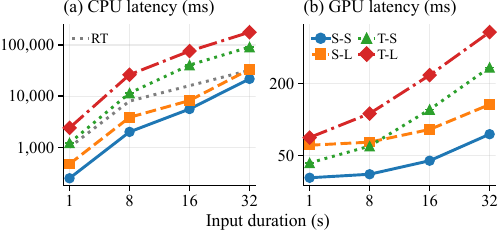}
\vspace{-20pt}
\caption{Latency versus input duration on one host: (a) one CPU thread, and (b) RTX 5090. S-S: SEAL (small); S-L: SEAL (large); T-S: TIGER (small); T-L: TIGER (large); RT: real-time boundary. Error bars (\textpm{} 1 s.d.) are smaller than the markers.}
\label{fig:latency}
\end{figure}

Under TIGER's protocol, SEAL (small) reaches 12.89~dB SI-SDRi and 13.62~dB SDRi at the lowest parameter and MAC cost in Table~\ref{tab:external}; against TIGER (small) it gains 0.31~dB SI-SDRi and 0.47~dB SDRi with 28\% fewer parameters and 2.9 times fewer MACs. MACs do not determine runtime, so Fig.~\ref{fig:latency} reports measured latency: for a 1~s input, SEAL (small) needs 32.7 vs.\ 43.5~ms on GPU and 247 vs.\ 1,211~ms on one CPU thread against TIGER (small). SEAL GPU latency is nearly flat up to 8~s and rises beyond 16~s with the quadratic temporal attention, yet both SEAL models stay below their TIGER counterparts through 32~s on CPU and GPU.

SEAL (large) doubles the shared refinement with the same cell and training recipe; its 0.59~M parameters and 4.85~G MACs remain 28\% and 37\% below TIGER (small). It reaches 13.66~dB SI-SDRi and 14.42~dB SDRi, 1.08 and 1.27~dB over TIGER (small), and exceeds TF-GridNet on both metrics at 67 times fewer MACs. It trails TIGER (large) by 0.07~dB SI-SDRi and leads it by 0.20~dB SDRi at 3.1 times fewer MACs.

\begin{table}[t]
\caption{Ablation study of SEAL (small) on the EchoSet test split. $\Delta$ is the signed SI-SDRi gap relative to Full. Bold: best SI-SDRi.}
\vspace{5pt}
\label{tab:ablation}
\centering
\small
\setlength{\tabcolsep}{0pt}
\begin{tabular*}{\columnwidth}{@{\extracolsep{\fill}}lcccc@{}}
\toprule
Configuration & Params (k) & MACs (G) & SI-SDRi & $\Delta$ \\
\midrule
Full SEAL              & 590.5 & 2.637 & \textbf{12.89} & --- \\
\midrule
\multicolumn{5}{@{}l}{\footnotesize\textit{Additive reconstruction controls}} \\
\hspace{6pt}w/o additive residual        & 589.5 & 2.620 & 11.85 & \textminus{}1.04 \\
\hspace{6pt}w/ single atom ($M\!=\!1$)   & 589.2 & 2.617 & 12.34 & \textminus{}0.55 \\
\midrule
\multicolumn{5}{@{}l}{\footnotesize\textit{Sparse expert routing controls}} \\
\hspace{6pt}w/ dense readout             & 363.2 & 2.539 & 11.99 & \textminus{}0.90 \\
\hspace{6pt}w/ fixed-hash routing        & 590.5 & 2.637 & 12.57 & \textminus{}0.32 \\
\hspace{6pt}w/o delta/temporal statistics & 590.5 & 2.637 & 12.10 & \textminus{}0.79 \\
\hspace{6pt}w/o additive step cue        & 590.5 & 2.637 & 12.59 & \textminus{}0.30 \\
\hspace{6pt}w/o norm cap ($\rho\!=\!\infty$) & 590.5 & 2.637 & 12.57 & \textminus{}0.32 \\
\bottomrule
\end{tabular*}
\end{table}

Table~\ref{tab:ablation} ablates each component with retrained controls of SEAL (small); all remaining results use this model. The fixed-checkpoint interventions below use validation and measure how much the trained model depends on a component, which need not equal the retrained gap.

\subsection{Reconstruction Accuracy and Cancellation}

Removing the additive residual lowers test SI-SDRi by 1.04~dB, saving 990 parameters (Table~\ref{tab:ablation}); this control keeps the mask correction in \eqref{eq:mask-correction}. The single-atom variant is 0.55~dB below Full; since one atom already reaches every point of the real allocation simplex, the gap reflects the factorized parameterization, not reachability.

To check that the model relies on the residual where components cancel, we compared $A_iX$ and $A_iX+R_i$ in the Full checkpoint on all 4,604 validation utterances, measuring bin-level cancellation by $|X|/(|S_1|+|S_2|+|N|+\epsilon)$, where $S_i$ is the reference STFT, $N=X-S_1-S_2$, and $\epsilon$ is a floor. Heavily cancelled bins fall below 0.325 (the 10th percentile of validation bins with summed speech magnitude $> -60$~dB relative to utterance maximum, and a smaller-to-larger source magnitude ratio $\ge 0.25$). Utterance NMSE divides the speech error power over these bins by their target power.

Adding the residual reduces mean NMSE from 0.891 to 0.552, a paired absolute reduction of 0.339 (95\% CI [0.336, 0.343]), or 38.0\% relative, so the model relies on the residual on cancelled bins. Disabling it costs 1.28~dB validation SI-SDRi (95\% CI [1.26, 1.30]), against the retrained gap of 1.04~dB.

\subsection{Conditional Capacity and Expert Assignments}
\label{sec:capacity-results}

Two controls separate what the expert bank stores from how tokens are assigned. The dense readout retains one always-on expert and the shared projection; the fixed-hash control keeps the six-expert bank and Top-1 dispatch but replaces learned assignment with a deterministic function of token coordinates, at Full's cost without router regularizers. They reach 11.99 and 12.57~dB test SI-SDRi against 12.89~dB for Full, so hashed dispatch adds 0.58~dB over a single expert, and learned assignment a further 0.32~dB at identical cost. Over the dense readout, Full adds 0.23~M parameters but only 3.7\% of its MACs.

Assignments also matter inside the network: permuting token--expert matches within each utterance and step, with expert counts, weights and correction strengths fixed, lowers mean validation SI-SDRi by 0.64~dB over 20 permutations (95\% CI [0.61, 0.67]). The same protocol costs 0.01~dB on the fixed-hash checkpoint (95\% CI [\textminus{}0.002, 0.022]), so the drop in Full reflects which expert each token receives rather than reassignment itself, and the 0.58~dB of the fixed-hash path does not come from emergent specialization.

\subsection{Refinement Evidence and Step-Cue Influence}
\label{sec:cue-results}
Disabling the inter-step delta and temporal-statistics branches of the routing query gives 12.10~dB ($\Delta=-0.79$~dB), their joint contribution; in this control (w/o delta/temporal statistics), $\beta_s$ still receives the delta summary. Removing the additive step cue costs 0.30~dB. Both controls keep learned assignment.

Without the norm cap the step cue can dominate the query; with it, the cue is confined to low-margin decisions. In the w/o norm cap control (150 validation utterances), the unconstrained step cue reaches 1.7--1.9 times the mean norm of the fused evidence at the first two steps, and removing it changes 85.5\% of first-step assignments; this control is 0.32~dB below Full. Projecting its cue onto radii from 0.05 to 0.60 and comparing the assignments with the unprojected ones, a radius of 0.15 changes 8.4--12.0\% of assignments per step. We fixed $\rho=0.15$ from this flip statistic without using separation quality. In the Full checkpoint, the cap is active and the bound of \eqref{eq:clip} holds: all four raw cue norms (4.37--4.95) are projected to 0.15, and zeroing the cue changes 33.6--44.8\% of assignments in the lowest quartile of the non-embedding Top-2 routing margin but at most 0.01\% in the highest.

\section{CONCLUSION AND FUTURE WORK}

We presented SEAL, a compact time--frequency separator that combines mixture-closed additive reconstruction with refinement-aware sparse routing in a shared cell. The additive residual relaxes the per-bin scaling limit while preserving the mixture sum, whereas the router selects token-specific expert corrections from acoustic and inter-step evidence, with a norm cap restricting the step cue to low-margin decisions. On EchoSet, SEAL (small) outperforms the reported TIGER (small) scores on both metrics with fewer parameters and MACs, and both SEAL configurations achieve lower measured latency than their TIGER counterparts. Retrained ablations confirm that both proposed mechanisms contribute to separation quality (Table~\ref{tab:ablation}). Our evaluation is limited to EchoSet and two-speaker mixtures; future work will examine additional corpora and speaker counts and replace quadratic temporal attention to reduce GPU latency for inputs longer than 16~s.

\clearpage

\bibliographystyle{IEEEbib}
\bibliography{refs}

@inproceedings{rong2024gtcrn,
  author = {Xiaobin Rong and Tianchi Sun and Xu Zhang and Yuxiang Hu and Changbao Zhu and Jing Lu},
  title = {{GTCRN: A Speech Enhancement Model Requiring Ultralow Computational Resources}},
  booktitle = {Proc. IEEE ICASSP},
  year = {2024}
}

@inproceedings{hershey2016dc,
  author = {John R. Hershey and Zhuo Chen and Le Roux, Jonathan and Shinji Watanabe},
  title = {{Deep Clustering: Discriminative Embeddings for Segmentation and Separation}},
  booktitle = {Proc. IEEE ICASSP},
  year = {2016}
}

@inproceedings{yu2017pit,
  author = {Dong Yu and Morten Kolb{\ae}k and Zheng-Hua Tan and Jesper Jensen},
  title = {{Permutation Invariant Training of Deep Models for Speaker-Independent Multi-Talker Speech Separation}},
  booktitle = {Proc. IEEE ICASSP},
  year = {2017}
}

@article{kolbaek2017upit,
  author = {Morten Kolb{\ae}k and Dong Yu and Zheng-Hua Tan and Jesper Jensen},
  title = {{Multitalker Speech Separation With Utterance-Level Permutation Invariant Training of Deep Recurrent Neural Networks}},
  journal = {IEEE/ACM Trans. Audio, Speech, Lang. Process.},
  volume = {25},
  pages = {1901--1913},
  number = {10},
  year = {2017}
}

@inproceedings{wang2023tfgrid,
  author = {Zhong-Qiu Wang and Samuele Cornell and Shukjae Choi and Younglo Lee and Byeong-Yeol Kim and Shinji Watanabe},
  title = {{TF-GridNet: Making Time-Frequency Domain Models Great Again for Monaural Speaker Separation}},
  booktitle = {Proc. IEEE ICASSP},
  year = {2023}
}

@article{bsrnn,
  author = {Yi Luo and Jianwei Yu},
  title = {{Music Source Separation With Band-Split RNN}},
  journal = {IEEE/ACM Trans. Audio, Speech, Lang. Process.},
  volume = {31},
  pages = {1893--1901},
  year = {2023}
}

@inproceedings{subakan2021sepformer,
  author = {Cem Subakan and Mirco Ravanelli and Samuele Cornell and Mirko Bronzi and Jianyuan Zhong},
  title = {{Attention Is All You Need in Speech Separation}},
  booktitle = {Proc. IEEE ICASSP},
  year = {2021}
}

@article{convtasnet,
  author = {Yi Luo and Nima Mesgarani},
  title = {{Conv-TasNet: Surpassing Ideal Time-Frequency Magnitude Masking for Speech Separation}},
  journal = {IEEE/ACM Trans. Audio, Speech, Lang. Process.},
  volume = {27},
  pages = {1256--1266},
  number = {8},
  year = {2019}
}

@inproceedings{dprnn,
  author = {Yi Luo and Zhuo Chen and Takuya Yoshioka},
  title = {{Dual-Path RNN: Efficient Long Sequence Modeling for Time-Domain Single-Channel Speech Separation}},
  booktitle = {Proc. IEEE ICASSP},
  year = {2020}
}

@inproceedings{sudormrf,
  author = {Efthymios Tzinis and Zhepei Wang and Paris Smaragdis},
  title = {{Sudo rm -rf: Efficient Networks for Universal Audio Source Separation}},
  booktitle = {Proc. IEEE MLSP},
  year = {2020}
}

@inproceedings{afrcnn,
  author = {Xiaolin Hu and Kai Li and Weiyi Zhang and Yi Luo and Jean-Marie Lemercier and Timo Gerkmann},
  title = {{Speech Separation Using an Asynchronous Fully Recurrent Convolutional Neural Network}},
  booktitle = {Advances in Neural Information Processing Systems},
  volume = {34},
  year = {2021}
}

@inproceedings{tdanet,
  author = {Kai Li and Runxuan Yang and Xiaolin Hu},
  title = {{An Efficient Encoder-Decoder Architecture with Top-Down Attention for Speech Separation}},
  booktitle = {Proc. ICLR},
  year = {2023}
}

@inproceedings{luo2021groupcomm,
  author = {Yi Luo and Cong Han and Nima Mesgarani},
  title = {{Ultra-Lightweight Speech Separation via Group Communication}},
  booktitle = {Proc. IEEE ICASSP},
  year = {2021}
}

@inproceedings{shazeer2017moe,
  author = {Noam Shazeer and Azalia Mirhoseini and Krzysztof Maziarz and Andy Davis and Quoc Le and Geoffrey Hinton and Jeff Dean},
  title = {{Outrageously Large Neural Networks: The Sparsely-Gated Mixture-of-Experts Layer}},
  booktitle = {Proc. ICLR},
  year = {2017}
}

@article{fedus2022switch,
  author = {William Fedus and Barret Zoph and Noam Shazeer},
  title = {{Switch Transformers: Scaling to Trillion Parameter Models with Simple and Efficient Sparsity}},
  journal = {J. Mach. Learn. Res.},
  volume = {23},
  pages = {1--39},
  number = {120},
  year = {2022}
}

@inproceedings{hu2026tfmoe,
  author = {Qinzhe Hu and Chenda Li and Wangyou Zhang and Shujie Liu and Yan Lu and Yanmin Qian},
  title = {{TF-MoE: Time-Frequency Mixture-of-Experts for Efficient Speech Separation}},
  booktitle = {Proc. Interspeech},
    year = {2026}
}

@misc{wang2022moess,
  author = {Xiaofei Wang and Zhuo Chen and Yu Shi and Jian Wu and Naoyuki Kanda and Takuya Yoshioka},
  title = {{Handling Trade-Offs in Speech Separation with Sparsely-Gated Mixture of Experts}},
  howpublished = {arXiv:2211.06493},
  year = {2022}
}

@inproceedings{bralios2023latent,
  author = {Dimitrios Bralios and Efthymios Tzinis and Gordon Wichern and Paris Smaragdis and Le Roux, Jonathan},
  title = {{Latent Iterative Refinement for Modular Source Separation}},
  booktitle = {Proc. IEEE ICASSP},
  year = {2023}
}

@inproceedings{wisdom2019consistency,
  author = {Scott Wisdom and John R. Hershey and Kevin Wilson and Jeremy Thorpe and Michael Chinen and Brian Patton and Rif A. Saurous},
  title = {{Differentiable Consistency Constraints for Improved Deep Speech Enhancement}},
  booktitle = {Proc. IEEE ICASSP},
  year = {2019}
}

@article{williamson2016crm,
  author = {Donald S. Williamson and Yuxuan Wang and DeLiang Wang},
  title = {{Complex Ratio Masking for Monaural Speech Separation}},
  journal = {IEEE/ACM Trans. Audio, Speech, Lang. Process.},
  volume = {24},
  pages = {483--492},
  number = {3},
  year = {2016}
}

@inproceedings{xu2025tiger,
  author = {Mohan Xu and Kai Li and Guo Chen and Xiaolin Hu},
  title = {{TIGER: Time-frequency Interleaved Gain Extraction and Reconstruction for Efficient Speech Separation}},
  booktitle = {Proc. ICLR},
  year = {2025}
}

@article{vincent2006bss,
  author = {Emmanuel Vincent and R{\'e}mi Gribonval and C{\'e}dric F{\'e}votte},
  title = {{Performance Measurement in Blind Audio Source Separation}},
  journal = {IEEE Trans. Audio, Speech, Lang. Process.},
  volume = {14},
  pages = {1462--1469},
  number = {4},
  year = {2006}
}

@inproceedings{leroux2019sdr,
  author = {Le Roux, Jonathan and Scott Wisdom and Hakan Erdogan and John R. Hershey},
  title = {{SDR --- Half-Baked or Well Done?}},
  booktitle = {Proc. IEEE ICASSP},
  year = {2019}
}

@inproceedings{scheibler2022sdr,
  author = {Robin Scheibler},
  title = {{SDR --- Medium Rare with Fast Computations}},
  booktitle = {Proc. IEEE ICASSP},
  year = {2022}
}

@inproceedings{zhao2023d2former,
  author = {Shengkui Zhao and Bin Ma},
  title = {{D2Former: A Fully Complex Dual-Path Dual-Decoder Conformer Network using Joint Complex Masking and Complex Spectral Mapping for Monaural Speech Enhancement}},
  booktitle = {Proc. IEEE ICASSP},
  year = {2023}
}

@inproceedings{zhang2024nass,
  author = {Zizheng Zhang and Chen Chen and Hsin-Hung Chen and Xiang Liu and Yuchen Hu and Eng Siong Chng},
  title = {{Noise-Aware Speech Separation with Contrastive Learning}},
  booktitle = {Proc. IEEE ICASSP},
  year = {2024}
}

@inproceedings{csordas2024moeut,
  author = {R{\'o}bert Csord{\'a}s and Kazuki Irie and J{\"u}rgen Schmidhuber and Christopher Potts and Christopher D. Manning},
  title = {{MoEUT: Mixture-of-Experts Universal Transformers}},
  booktitle = {Advances in Neural Information Processing Systems},
  volume = {37},
  year = {2024}
}

@inproceedings{qiu2025rmoe,
  author = {Zihan Qiu and Zeyu Huang and Shuang Cheng and Yizhi Zhou and Zili Wang and Ivan Titov and Jie Fu},
  title = {{Layerwise Recurrent Router for Mixture-of-Experts}},
  booktitle = {Proc. ICLR},
  year = {2025}
}

@inproceedings{chen2022soundspaces2,
  author = {Changan Chen and Carl Schissler and Sanchit Garg and Philip Kobernik and Alexander Clegg and Paul Calamia and Dhruv Batra and Philip W. Robinson and Kristen Grauman},
  title = {{SoundSpaces 2.0: A Simulation Platform for Visual-Acoustic Learning}},
  booktitle = {Advances in Neural Information Processing Systems (Datasets and Benchmarks Track)},
  volume = {35},
  year = {2022}
}

@misc{zoph2022stmoe,
  author = {Barret Zoph and Irwan Bello and Sameer Kumar and Nan Du and Yanping Huang and Jeff Dean and Noam Shazeer and William Fedus},
  title = {{ST-MoE: Designing Stable and Transferable Sparse Expert Models}},
  howpublished = {arXiv:2202.08906},
  year = {2022}
}
\end{document}